\documentclass[authorversion, nonacm]{acmart}

\AtBeginDocument{%
  \providecommand\BibTeX{{%
    \normalfont B\kern-0.5em{\scshape i\kern-0.25em b}\kern-0.8em\TeX}}}

\usepackage{fancyhdr}

\begin{document}

\title{My Future with My Chatbot: A Scenario-Driven, User-Centric Approach to Anticipating AI Impacts}

\author{Kimon Kieslich}
\email{k.kieslich@uva.nl}
\orcid{0000-0002-6305-2997}
\author{Natali Helberger}
\email{n.helberger@uva.nl}
\orcid{0000-0003-1652-0580}
\affiliation{%
  \institution{University of Amsterdam}
  \streetaddress{Postbus 15514}
  \city{Amsterdam}
  \country{The Netherlands}
  \postcode{1001 NA}
}

\author{Nicholas Diakopoulos}
\email{nad@northwestern.edu}
\orcid{0000-0001-5005-6123}
\affiliation{%
  \institution{Northwestern University}
  \streetaddress{70 Arts Circle Drive}
  \city{Evanston, IL 60208}
  \country{USA}}

\setcopyright{none}

\renewcommand{\shortauthors}{Kieslich, Helberger, \& Diakopoulos}
\renewcommand{\shorttitle}{My Future with My Chatbot}
\begin{abstract}
As a general purpose technology without a concrete pre-defined purpose, personal chatbots can be used for a whole range of objectives, depending on the personal needs, contexts, and tasks of an individual, and so potentially impact a variety of values, people, and social contexts. Traditional methods of risk assessment are confronted with several challenges: the lack of a clearly defined technology purpose, the lack of clearly defined values to orient on, the heterogeneity of uses, and the difficulty of actively engaging citizens themselves in anticipating impacts from the perspective of their individual lived realities. In this article, we leverage  scenario writing at scale as a method for anticipating AI impact that is responsive to these challenges. The advantages of the scenario method are its ability to engage individual users and stimulate them to consider how chatbots are likely to affect their reality and so collect different impact scenarios depending on the cultural and societal embedding of a heterogeneous citizenship. Empirically, we tasked 106 US-based participants to write short fictional stories about the future impact (whether desirable or undesirable) of AI-based personal chatbots on individuals and society and, in addition, ask respondents to explain why these impacts are important and how they relate to their values. In the analysis process, we map those impacts and analyze them in relation to socio-demographic as well as AI-related attitudes of the scenario writers. We show that our method is effective in (1) identifying and mapping desirable and undesirable impacts of AI-based personal chatbots, (2) setting these impacts in relation to values that are important for individuals, and (3) detecting socio-demographic and AI-attitude related differences of impact anticipation.
\end{abstract}

\ccsdesc[500]{Applied computing~Law, social and behavioral sciences}
\ccsdesc[500]{Human-centered computing~Empirical studies in HCI}
\ccsdesc[300]{Social and professional topics~Government technology policy}
\ccsdesc[300]{Social and professional topics~User characteristics}

\keywords{chatbots, anticipatory governance, scenario-writing, impact assessment}

\pagestyle{fancy}
\fancyfoot[R]{%
  \ifnum\value{page}>1 
    To be published at ACM FAccT'24 
  \fi
}



\maketitle

\hypertarget{introduction}{%
\section{Introduction}\label{introduction}}

Generative AI systems, such as ChatGPT, have considerably lowered the threshold for millions of users to employ very powerful AI technology in their daily lives. From the release of ChatGPT as a service, it took only a year before Open AI announced that people could build their own personal chatbot (GPT) to assist them in their personal and professional lives. Given the proliferation of the technology, not only through OpenAI, but also as it is integrated into other products and made available through open source platforms, what will be the impact on individuals and society? As a general purpose technology without a concrete pre-defined purpose, personal chatbots could be used for a whole range of objectives, depending on the personal needs, contexts, and tasks of an individual, and in doing so potentially impact a variety of values, people, and social contexts. This creates challenges for traditional methods of risk assessment including  the lack of a clearly defined technology purpose, the lack of a clearly defined values to orient on, the heterogeneity of uses, and the difficulty of actively engaging citizens themselves in anticipating impacts from the perspective of their individual lived realities \cite{the_danish_institute_for_human_rights_human_2020}. Additionally, these approaches are mostly based on domain experts' knowledge and opinions (e.g., technology developers; researchers) or reviews of the existing scientific literature to identify and classify potential risks of an emerging technology \cite{bucinca_aha_2023, shelby_sociotechnical_2023, solaiman_evaluating_2023, weidinger_sociotechnical_2023}. However, relying solely on domain experts might be too shortsighted, as experts also carry biases and thus might fail to anticipate future impacts for specific stakeholder groups and/or society \cite{nanayakkara_unpacking_2021}. Research has also shown that inventors (developers) of AI systems tend to be overly optimistic about the positive features of it, even if it is comparably low-performing \cite{cratsley_inventors_2023}, which makes a critical assessment of the technology even more unlikely. 

In this paper, we advance scenario writing at scale as an alternative method for anticipating AI impact \cite{kieslich_anticipating_2023}. The advantages of the scenario method is its ability to engage individual users and stimulate them to consider how personal chatbots are likely to affect their reality and collect different impact scenarios depending on the cultural and societal embedding of a heterogeneous citizenship. In doing so, we also answer the call of various risk assessment frameworks that highlight the need of engaging affected people and communities in impact assessment \cite{mantelero_evidence-based_2021, mesmer_auditing_2023, stahl_systematic_2023, the_danish_institute_for_human_rights_human_2020}. Furthermore, we also connect these anticipations to respondents’ values, i.e. we elaborate on the question of why specific outlined impacts are deemed as important. An empirical assessment is of particular relevance considering the value-laden approaches to ethical or responsible AI formulated by policy-makers, tech corporations and civil society \cite{jobin_global_2019}. 

We define chatbots as conversational agents that, based on natural-language processing technology, interact with users in human-like conversations to provide access to data or information or fulfill services for them \cite{folstad_chatbots_2017, bodrunova_different_2019, jeon_beyond_2023}. Chatbots can be both text-based or react to voice commands and are already used in a multitude of application domains \cite{bodrunova_different_2019, jones_public_2019}. General purpose chatbots like ChatGPT are of particular interest as they are user-driven and can, in principle, generate answers to a wide range of human inputs that are only subject to the system’s own safety moderation protocols. LLM-based chatbots can also build long-term relations with users as they can incorporate the historical context of interaction from specific users. Several researchers have already warned about the ethical challenges that LLMs that engage in interaction with lay people can entail \cite{weidinger_ethical_2021, weidinger_sociotechnical_2023, weidinger_taxonomy_nodate}. 

In this study, we use large-scale scenario-writing with more than a hundred participants to collect unique future foresight of participants about the impact of LLM-based personal chatbots. We detect a wide range of both desirable and undesirable future impacts of this technology and distinguish between individual and societal impacts. We enrich our impact classification with respondents’ value beliefs that help in explaining why impacts matter to them. Further, we not only highlight the frequently mentioned impact themes, but also shed light on the long-tail of impacts that might only be of importance for sub-groups of the population. Additionally, we quantify the presence of impacts to analyze how various background variables from participants are related. Our findings, thus, contribute to the current work on AI impact assessment by introducing citizens’ anticipations of future impact to the discourse. 

\hypertarget{related-work}{%
\section{Related Work}\label{related-work}}

In this section we first draw on the literature on anticipatory governance and participatory foresight to ground our scenario-writing approach and explain its added value to the existing AI impact assessment literature. We also describe the strengths of scenario-writing as a method to identify impacts. Then, we outline how human values are currently addressed in AI development and governance approaches. We highlight the importance of including citizens’ perceptions and value preferences in this process and reflect on the current literature on public opinion on ethical AI principles.

\hypertarget{governance}{%
\subsection{Anticipatory Governance and Participatory Foresight for Impact Assessment}\label{governance}}

Political governance and regulation approaches, but also research in social sciences oftentimes struggle because of the pace of technological progress, the complexity of social and economic dynamics as well as the difficulty to anticipate the actual impact of a technology like chatbots on society, and society's values. The need for anticipatory methods to identify potential risks is, for example, highlighted by the EU AI Act with its focus on ‘reasonably  foreseeable risks’ \cite{european_parliament_texts_2023}. Anticipating how AI will develop and impact society can inform regulatory and risk mitigation strategies but also strategic innovation decisions, responsible development \cite{guston_understanding_2013, selbst_institutional_2021}, and wider governance strategies, such as investments in research, education, or empowering affected stakeholders or issuing implementation acts and (safety) guidelines \cite{helberger_futurenewscorp_2024}. 

Anticipating societal impacts in the early  development and deployment phase allows risk management at an early stage and, ideally, prevents harmful societal impacts because they materialize or scale, and can make early interventions more effective and less costly \cite{cave_bridging_2019, diakopoulos_anticipating_2021, fuerth_operationalizing_2011, guston_understanding_2013}. Importantly, because it is impossible to predict the future with any certainty \cite{helberger_futurenewscorp_2024}, anticipatory governance studies aim for showing a range of potential future developments \cite{quay_anticipatory_2010}. These visions can be combined with value-related expectations of affected stakeholders. Consequently, anticipatory governance approaches also tap into the question of how stakeholders want a technology to be developed \cite{guston_understanding_2013, mittelstadt_how_2015}. 

In connection with AI technology, several frameworks have been introduced, which aim to map the potential risks and impacts of AI on individuals and society \cite{moss_assembling_2021, stahl_systematic_2023}. The core function of AI impact assessment is to identify risks for human rights, ethical values, the environment or social conditions, and – in the best case – report and document them in an early stage so that countermeasures and mitigation can be undertaken, for instance through regulation \cite{selbst_institutional_2021} or re-design. AI impact assessments have been performed for several applications like LLMs \cite{weidinger_ethical_2021}, text to image technology \cite{bird_typology_2023}, and generative AI \cite{magooda_framework_2023, solaiman_evaluating_2023, weidinger_sociotechnical_2023}. Further, some researchers proposed more general techniques to assess impacts \cite{bucinca_aha_2023, hoffmann_adding_2023, shelby_sociotechnical_2023}. Impact assessment prepares developers, scholars, regulators and all affected stakeholders for potential pitfalls and includes not only technical risks, but also societal ones \cite{metcalf_algorithmic_2021, moss_assembling_2021}. In other words; following the understanding of AI as a socio-technical system, the potential risks from a general purpose technology, such as chatbots, are significantly determined by the way the technology is used by individuals in their respective social or organizational embedding \cite{ad_hoc_expert_group_outcome_2020}. This is also why the involvement of affected stakeholders is seen as an important element in impact assessments \cite{stahl_systematic_2023}. General purpose technologies, like generative AI or chatbots, can be potentially used for a whole range of objectives and in very different contexts. A key challenge for impact assessments is being able to identify which impacts to analyze in the first place \cite{ramirez_plausibility_2014}. Therefore, an important first step to be able to engage in a risk assessment is mapping and prioritizing different impact scenarios. 

Scenario-writing can be used as a method to facilitate anticipatory thinking and to develop several alternative visions of the future \cite{amer_review_2013, amosbinks_anticipatory_2023, borjeson_scenario_2006, selin_trust_2006}. The further away in the future, the higher the range of possible scenarios that are imaginable. The key task for future studies is to anticipate not only possible futures, but also plausible or even probable ones \cite{ramirez_plausibility_2014}. Future thinking is embedded in current development as current research priorities and policy measures can heavily influence how technology evolves in the future \cite{cave_bridging_2019}. Additionally, scenarios are embedded in a real-world setting \cite{amer_review_2013} and display vivid examples of potential future developments that are more relatable, which makes them accessible for non-experts like the broad public or policy-makers \cite{helberger_futurenewscorp_2024}. As such, future scenarios can help in facilitating a discussion about the desirability of different development paths, and what measures are needed to realize/prevent those scenarios \cite{stahl_virtual_2013}. Scenario-writing can serve as a risk-assessment tool \cite{helberger_futurenewscorp_2024, kieslich_anticipating_2023} as it helps to “identify potential and unexpected risks resulting from the deployment of AI into a particular socio-technical context, while engaging also non-experts and diverse communities” \cite{guston_understanding_2013}. Consequently, when it comes to the realization of scenario-writing it is crucial to consider who the responsible actors are that define and write scenarios in the first place \cite{metcalf_algorithmic_2021}. Relying on the future visions of the industry representatives or researchers alone risks missing out other relevant perspectives that are affected by and/or contribute to the realization of a particular future implementation of a technology \cite{barnett_crowdsourcing_2022, bonaccorsi_expert_2020}. Researchers thus acknowledge the value of non-experts to anticipate consequences for social and everyday life \cite{nikolova_rise_2014}. Moreover, from a normative viewpoint, society should be included in the decision-making process on how they want to engage with high-impact AI-systems \cite{calvo_advancing_2020, rahwan_society---loop_2018}. An important advantage of the scenario method is that it can be more engaging and also stimulate non-experts to anticipate the impacts that a particular technology has on their individual societal context \cite{european_digital_rights_artificial_2021, heidingsfelder_expanding_2016, schuijer_citizen_2021}. 

One approach to mitigate blind spots relates to participatory foresight \cite{brey_anticipatory_2012, nikolova_rise_2014}. The idea of participatory foresight is to engage a diverse set of participants in anticipating future impact. In this way potential blind spots can be filled out and pluralistic visions of the future emerge that reflect the realities and circumstances of all involved and, thus, to highlight lived experiences from a diverse group of actors as valuable source of knowledge \cite{birhane_power_2022}. While some researchers highlight the necessity to pursue this design \cite{metcalf_algorithmic_2021, moss_assembling_2021}, only some empirical studies \cite{barnett_crowdsourcing_2022, diakopoulos_anticipating_2021, kieslich_anticipating_2023} have been conducted that utilize this approach in the context of AI. By combining participatory foresight and scenario methods it is possible to not only investigate individual impacts but understand through this more qualitative approach also the different socio-technical contexts in which a technology can potentially operate, as well as values and value-trade offs that are potentially involved. 

\hypertarget{values}{%
\subsection{Values and AI}\label{values}}

Values play an important role as a benchmark in deciding whether a particular future scenario is desirable or not. Meynhardt explains: “As natural response, humans feel positive about something if there is a direct personal gain or a positive impact on the community or society they live in. This is what public value refers to” \cite{meynhardt_public_2019}. A future application of AI that conflicts with individual expectations of privacy or autonomy or societal values such as equality, political freedom or solidarity is potentially less desirable than an application that respects those values, unless there is a trade-off with other values that participants may find even more important. In general, values reflect a normative notion of what is good or bad \cite{gabriel_artificial_2020, riedl_using_2016, van_de_poel_embedding_2020}. Values can be understood both as a form of normative ideal, but can also refer to people’s beliefs \cite{van_de_poel_embedding_2020}. People can value different things, their values must not necessarily be normatively good, and values can also conflict and require trade-offs \cite{dignum_responsible_2019, van_de_poel_embedding_2020}. In general, “values serve as guiding principles of what people consider important in life” \cite{cheng_developing_2010} and  “are determinants of virtually all kinds of behavior that could be called social behavior or social action, attitudes and ideology, evaluations, moral judgments and justifications of self to others, and attempts to influence others” \cite{rokeach_nature_1973}. In addition, values can also have a societal function as they “help communities of individuals resolve collective-action problems, stabilise social relationships, and flourish over time” \cite{gabriel_artificial_2020}. 

Values can offer  guidance on how AI should be developed and implemented, thus, enabling them a priori to follow ‘good’ principles \cite{nissenbaum_values_1998, umbrello_mapping_2021, van_de_poel_embedding_2020}. Values also guide public and private actors in their decision making, including in their strategy to new technologies \cite{andrews_public_2019}. For example, the European Commission \cite{european_commission_ethics_2019}, in their ethics approach to trustworthy AI, follows a value-based approach that should pave the way for an ethically sound implementation of AI, and under the AI Act, the potential to cause significant harm to fundamental rights or public values, such as the environment, democracy or the rule of law, can be a reason for the qualification as a high-risk AI system \cite{european_parliament_texts_2023}. But the EU is not alone: numerous other organizations developed ethical guidelines (that are partially based on values) on how to tackle the ramifications of AI \cite{hagendorff_ethics_2020, jobin_global_2019}. In particular, these frameworks treat value-orientation in the sense that their objective is to assess the impact of AI, or particular AI applications on values such as ethical and professional values, or human rights. However, many scholars criticize that these guidelines are toothless and are developed top-down without the consultation of the “society” itself \cite{hagendorff_ethics_2020, zuger_ai_2023}, or that society is not consulted on the operationalisation of - often vaguely formulated - abstract values. While there is much focus in the technical literature on the challenge of “value alignment” for AI systems \cite{christian_alignment_2020}, gaping questions remain about whose values should be aligned to, and why. 

In consideration of the critique of top-down approaches, Gabriel formulated two main difficulties for this task: “The first is to specify what values or principles AI should align with. [...] The second major difficulty concerns the individual or body of people who select the principles with which AI aligns.” \cite{gabriel_artificial_2020}. This is a crucial point given that some groups prioritize ethical values with respect to responsible AI differently \cite{jakesch_how_2022}. These difficulties are  also in line with the call of Gordon et al., who stress, in reference to AI governance, the importance of researching “how values manifest and ‘map’ among context-sensitive computational and social processes in the first place” \cite{masso_basic_2023}. Further, Masso et al. highlight that “a better understanding of the citizens’ perspectives of the values [...] would also contribute to preventing the potential negative behavioural impacts, like harms and risks, when designing, using, and implementing AI systems” \cite{masso_basic_2023}. Thus, there is a need to assess citizens’ values and perceptions regarding AI development and enrich the current top-down approaches with more diverse user perspectives. 


Given the expediency of including diverse value perspectives it is also necessary to research which sociodemographic and AI related characteristics are related to the development of different perceptions. Empirical studies have found that perceptions of AI (incl. moral trade-offs and value preferences) can vary with respect to gender, age, educational level, income and self-reported knowledge \cite{carradore_peoples_2022, fietta_dissociation_2022, ikkatai_octagon_2022, kieslich_ever_2023, zhang_artificial_2019, kieslich_artificial_2022, masso_basic_2023}. More specifically, men, more highly educated, younger, and having a higher income are attributes associated with a more positive view on AI – or, in short, perceptions of AI are more optimistic among those groups in the population, who are expected to profit the most from the technology. On the other hand, that does not imply that their counterparts are critical towards AI. Survey studies also showed that especially those groups who might be most affected by negative impacts of technology (e.g., people with low income; ethnic minority groups) are the ones who most likely do not engage with AI at all \cite{bao_whose_2022, kieslich_artificial_2022, kieslich_ever_2023}. The low level of awareness of ethical AI issues among vulnerable groups, coupled with an already difficult structural access to public participation, means that these groups are rarely heard in the discourse on the social consequences of AI.

This study is based on the hypothesis that scenario writing can be a tool to access and prompt for values that are otherwise hard to measure directly \cite{dignum_responsible_2019}. Thus, we tasked respondents to create a scenario and to reflect on their values and motivations that lead them to outline the specific impacts. We understand scenarios as a form of narrative symbolizing a cultural artifact that mirrors people’s beliefs on how society works or ought to work. Thus, in asking respondents to write a story outlining an impact, we implicitly learn something about what values matter to them, what values they see at risk, or where they see value trade-offs. While many of the referenced studies rely on quantitative survey data with standardized items, in this study, we focus on open ended answer collection, i.e. in the form of collecting scenarios, and elaborate on the differences between socio-demographic information and AI-related attitudes. 


\hypertarget{method}{%
\section{Method}\label{method}}

\hypertarget{procedure}{%
\subsection{Procedure}\label{procedure}}

To gauge the imaginations of (potential) users of personal chatbots, we deployed an online survey with an integrated scenario-writing task. The questionnaire was structured as follows: After respondents’ were introduced to the study’s goal and gave informed consent to participate in the study, they gave some information about their sociodemographic background, including information about their gender, race, educational level, income, and whether they had experienced discrimination. Then, respondents answered items about their AI related attitudes. We measured interest in AI with four items on a five point scale \cite{dosenovic_methodensteckbrief_2022} (mean index, Cronbachs \(\alpha\)=.91; M=3.09; SD=0.92), media consumption of AI related news with four items on a six-point scale \cite{dosenovic_methodensteckbrief_2022} (maximum score, M=4.06; SD=1.26), and personal use of AI-based chatbots with a self-developed item on a six point scale. 

Afterwards, respondents were introduced to the scenario-writing and background information about the technology. We defined a scenario as a “short fictional story that includes: (1) a setting of time and place, (2) characters with particular motivations and goals, and (3) a plot that includes character actions and events that lead to some impact of interest.” As technological background information, we described capabilities (prompt examples, a range of functionalities, general purpose applicability, personalization) as well as limitations (accuracy, privacy \& security, biases, lack of contextual understanding) of AI-based personal chatbots and provided a screenshot of the start page of ChatGPT. We instructed respondents that their scenarios should be creative, believable and plausible and promised an extra payment of \$2 USD for the top ten percent of the scenarios evaluated by the authors of this paper. This approach was adapted from \cite{diakopoulos_anticipating_2021, kieslich_anticipating_2023}. The full item wording, task setup and presentation can be found in Appendix 1-4. 

After respondents confirmed that they understood the instructions, we tasked them to write their scenario in an open text field. To discourage respondents’ use of ChatGPT to write the scenarios (which is an emerging issue in crowdwork \cite{veselovsky_artificial_2023}), we (1) instructed them explicitly not to use AI-assistance to write the scenario, (2) disabled copying and pasting in the text field, and (3) asked respondents to confirm that they did not use AI-assistance to write the scenario \cite{veselovsky_prevalence_2023}. To ensure an adequate scenario length, we set a minimum of 1,000 characters for the scenario which was enforced by the survey system such that participants could not submit until the scenario reached the minimum required length. After writing the scenario, we tasked respondents to think about the underlying values, which led them to outline the respective impacts in their scenarios. Concretely, we asked them: “Please explain why the impacts that you outlined in your scenario are important to you. Think also about how they relate to your ideas on how society/the world should be in terms of what you value. Please write at least 50 words!” Afterwards, respondents were debriefed and thanked. Participation in the study took on average 44.70 minutes (SD=29.85) and respondents were paid \$9.75 USD. 

\hypertarget{sample}{%
\subsection{Sample}\label{sample}}

 We recruited 135 respondents living in the US, who indicated that they speak English fluently, via the participant pool of the market research provider Prolific. Furthermore, we opted for a balanced gender distribution in the sample. Given the exploratory nature of this study, we didn’t apply other quota criteria to the sample. We applied several measures to ensure data quality in our sample. First, we filtered out all cases, in which scenario writers didn’t write about personal chatbots, but some form of other AI-based technology or totally unrelated topics. That resulted in an elimination of 15 cases. Second, although we applied methods to prevent the use of LLMs in the writing task, we checked the scenarios for being written by a LLM. We utilized GPTZero to assess if a scenario has been produced by a LLM \cite{tian_gptzero_2023}. We flagged all scenarios that had a probability score of 80 percent or more and contacted all corresponding respondents that received a flag. Out of the 16 contacted respondents, 14 did not counter-argue and withdrew their participation in the study when confronted with this information. This filtering resulted in a final sample of 106. There was a reasonable distribution of the final set of participants across the various sociodemographic variables we measured, including gender, race, and education (See Appendix 5 for full details).  

\hypertarget{analysis-methods}{%
\subsection{Analysis Methods}\label{data-filtering}}

The scenarios were analyzed with a qualitative thematic analysis approach \cite{glaser_discovery_2017, lofland_analyzing_2022}. We applied open and axial coding of scenarios to identify impacts. Excerpts from  the scenarios were used in a constant comparison approach to develop codes and establish emerging themes. With this approach, we were able to typologize and structure the content of the scenarios. Additionally, we used the responses from the value question to explore the rationale for why specific impacts were described. Once we had developed a taxonomy, we then quantified the occurrence of impact types for each respondent. For each scenario we identify the presence or absence of each impact type and use this to count the number of times each is mentioned across all scenarios. We then used descriptive statistics and logistic regression models to analyze this count data with respect to sociodemographic background variables.

\hypertarget{findings}{%
\section{Findings}\label{findings}}


\hypertarget{domains}{%
\subsection{Which domains are associated with the use of chatbots?}\label{domains}}

We first look at the usage domains in which chatbots were described in the scenarios, in other words, in which domains chatbots were most likely to have an impact, according to our panelists. The most prevalent use case mentioned was in people’s private everyday life (n=51). The most prevalent codes within this domain were the use as a personal assistant for scheduling, casual chats, or as a smart home device. Some scenarios also outlined specific uses, such as use as a life-coach, as a diet advisor, dating tools (either as advisor or as artificial companion), or for information retrieval (substitution for search engines). The second most prevalent use case was for business purposes (n=21). Chatbots were described as fulfilling a variety of functions, for instance, scheduling appointments, brainstorming, or editing texts and presentations. Some scenarios also described the use of chatbots in the recruitment process, either in helping users to compile CVs and motivation letters, or in the use of human resource teams to screen applicants. Further work-related application domains mentioned were marketing and customer service. 

Another occasionally mentioned domain was the use for creative tasks (n=15). Here, we distinguish between the use of chatbots for story-writing, entertainment, and art. In regard to story-writing (e.g., novels, narratives), chatbots were described as a brainstorming partner. The scenarios outlined how users chatted with chatbots to generate ideas or help with editing. The use in entertainment was connected to the entertainment industry, where scenarios described negative consequences for jobs. Additionally, two scenarios described the use of chatbots for the creative tasks of composing and inspiration for creating artworks. Some scenarios (n=9) also dealt with the use of chatbots for medical purposes. Characters used chatbots to get medical advice for physical problems (e.g., diagnosis of symptoms, or check-ups), but also for daily life advice (e.g., healthy diet). Another dimension in the medical domain is the support for mental health issues. Some scenarios described how people turned to chatbots in search of companionship and mental struggles. Finally, some scenarios also described that hospitals used chatbots to work more efficiently. A few scenarios dealt with the use of chatbots in education (n=6). Characters in the scenarios used chatbots to fulfill assignments for college or school with diverging motivation and effects. Some scenarios outlined a responsible and effective use of chatbots to foster educational goals, while others focused on negative aspects like cheating, or the loss of creativity. Sporadically mentioned use cases were the political domain (n=2), legal domain (n=3) and in science (n=1).   

\hypertarget{impacts}{%
\subsection{What are the impacts of chatbots?}\label{impacts}}

To analyze the impacts of chatbots we developed a classification scheme that we present in terms of two dimensions: desirable vs. undesirable and individual vs. societal. Furthermore, we explored respondents’ value statements to illuminate why specific impacts were articulated and how the scenarios related to the expected benefits or detriments for individuals or society. We included value statements as quotes in the following sections to contextualize and enrich the mention of impacts based on the moral beliefs and values of the scenario-writers. Desirable and undesirable impacts were chosen as the first dimension as we wanted to ground our analysis in participants’ current AI-related associations and not force them to artificially think about only negative impacts. Personal chatbots arguably can have both desirable and undesirable impacts and mapping and discussing what is on people’s mind is important to inform a debate about strategies on how to mitigate risks and amplify benefits. Furthermore, we take individual vs. societal level as an analytical lens because different values can be associated with those levels (e.g., concerns about societal development vs. personal well-being). Additionally, the scale and severity of risks and benefits vary between the levels. Lastly, the distinction can also help to identify spill-over effects from the individual sphere into the social sphere and vice versa.


\hypertarget{desirable-individual-impacts}{%
\subsubsection{Desirable Individual Impacts}\label{desirable-individual-impacts}}

A large proportion (n=47, 44.3\%) of the scenarios described desirable impacts of chatbots on the personal sphere of users’ lives. Sub-codes identified include: well-being, skills, positive performance of chatbots, and strengthening of human-human relationships.

The scenarios mention various ways in which chatbots can positively influence users’ well-being. We found some mentions of improvement of physical well-being, or helpful medical advice for physical impacts. Most impacts were related to mental well-being. That is, the users in the scenarios experienced encouragement (“AI can give you the confidence it requires to put yourself out there.” [P 89]) and confidence through communicating with personal chatbots. Specifically, chatbots were described as helping with work-life balance (“The AI in my story helped my main character find balance in her life.” [P 28]), losing weight, building safe spaces, and improving mental well-being through companionship. In terms of diversity, it is especially interesting to mention that some scenarios described how troubled persons (e.g., people suffering from depression) could especially benefit from using chatbots (“I believe they can provide a friend, an ear, someone or something that can listen and get people through their toughest of times.” [P 31]). 

Many scenarios described the positive performance of chatbots, for instance, users' work benefiting from the use of chatbots, students achieving learning goals much faster, or chatbots even helping for personal goals like increasing likes on dating platforms. Many of the benefits were based on efficiency of the chatbots (“Chatbots and AI can cut through the nonsense and provide efficient answers including step-by-step guidance in some instances. [P 4])” and their potential to make daily life easier. They could also help users in finding and condensing information, or automate tasks at work that were formerly time-intense. Moreover, another quality that was ascribed to chatbots was that they offered easier accessibility to information for users and that they made life more convenient. Further, some scenarios suggested a positive effect on skills, mainly in fostering knowledge improvement. That was based on the assumption that users’ could use chatbots to learn faster and in an easy to understand way. Some scenarios described that users sought chatbots to help them deal with social interactions, for instance in seeking advice on how to communicate with teenagers or giving advice for socially anxious people. On a work-related level, some scenarios described individual job opportunities due to the introduction of chatbots. These job opportunities arise out of a smart use of chatbots, for instance, in creating more creative and profound outputs at work that lead to promotions or bonuses (“AI-based chatbots can help individuals achieve their dreams and aspirations with a vast amount of ideas.” [P 80]). Some scenarios also described the gain of revenue for companies that, in turn, resulted in a profit for individuals, for instance, if they were owners of small businesses. Additionally, scenarios described that characters had more time to do non-work related tasks that, in turn, coincided with a better work-life balance.


\hypertarget{desirable-societal-impacts}{%
\subsubsection{Desirable Societal Impacts}\label{desirable-societal-impacts}}

Only five scenarios (4.7\%) outlined desirable impacts of the use of chatbots for society. Some scenarios set impacts in the context of societal trends and/or social cohesion. These few scenarios describe that the use of chatbots lead to strengthening collective action as it helps communities to work together on common goals (“The story emphasizes the positive role AI can play in fostering collective well-being and addressing societal challenges. It aligns with my vision of a world where technology serves as a tool for connection, empowerment, and positive change rather than a source of isolation or harm” [P 84]). Another positive impact that was mentioned was the promotion of multiculturalism as chatbots enable easier learning of languages and translations that contribute to better understanding among citizens. One scenario described that the large-scale use of chatbots could lead to a reduction in income equality as people were empowered and set equal in the workforce. Lastly, one scenario described chatbots as a tool that could be used to fight discrimination in the workplace, though it wasn’t specific in how it did so.

\hypertarget{undesirable-individual-impacts}{%
\subsubsection{Undesirable Individual Impacts}\label{undesirable-individual-impacts}}

A substantial number of scenarios outlined undesirable impacts on the individual level (n=40, 37.7\%). We identified sub-codes including: skills, well-being, worsening human-human relationships, negative performance of chatbots, misconduct, and over-reliance.

When scenarios dealt with skills, they highlighted the loss of creativity (“I value human experience and the fact that someone can't be defined by their resume or what they look like on paper.” [P 55]), loss of critical thinking (“I feel like the more we use AI, the more dumb we're all going to become, because we won't need to use our own brains anymore.” [P 2]), loss of literacy, or the loss of human interaction skills (“We humans need to connect with one another often and when we take away the human connection from most things the world starts to feel robotic and bland.” [P P18]). In this regard, chatbots were perceived as a replacement that, on the one hand, makes life and tasks easier, but on the other hand brings users to the point that they do not need to put in effort to achieve goals. Thus, the fear, human skills would become more meaningless and could vanish. 

Regarding well-being and affective reactions, we identified various sub-codes. Scenarios described users as being frustrated, worried, or ashamed as a consequence of using chatbots. For example, characters in the scenarios expressed frustration over poor-functioning chatbots, for instance, in customer service. Some scenarios also described negative impacts on well-being. These negative impacts were related to a loss of control of users or an over-dependency (“When the machines are placed into wrong positions, they may well end up punishing honest human work, incentivizing dishonesty and the embrace of a post-truth worldview.” [P 109]). In these cases, users got lost in communication with the chatbots and found themselves isolated in virtual communications. Concerning the influence of human-human relationships, some scenarios described – also related to the negative impact for well-being – a decline in human interactions. Due to the reliance on chatbots, human interaction is perceived to be diminished, including a loss of human-human communication. Some scenarios also outlined poor technical performance of chatbots. This could have multiple reasons, be it technical glitches, a lack of accuracy, or a loss of human touch that resulted in a negative evaluation of the chatbot's performance. Often, these negative performances were tied to specific application domains. One area that was deemed as especially negatively influenced was customer service, where chatbots were described as not helpful and leading to frustration for users. Another frequently anticipated negative aspect was associated with the use of chatbots in a work setting. This was mostly due to the fear of job losses due to the introduction of chatbots in workplace settings. The reasons for job loss could either be (1) overreliance of chatbots that lead characters to bad quality work (“Even if AI is knowledgeable, they can misunderstand us and send us in the wrong direction.” [P P14]) and get them fired, or (2) that their job was taken over by a chatbot. Furthermore, some scenario-writers also outlined potential misconduct with chatbots. This was expressed through characters working out of malicious intent, for instance, malicious users that deployed chatbots for spreading misinformation or cheating in exams or in application processes. Moreover, multiple scenarios outlined privacy violations due to the use of chatbots, i.e. that chatbots gained/accessed knowledge about sensible data and used it sometimes against the will of users (“I also think people should be careful of the information they send to artificial intelligence bots as that information ultimately belongs to someone else.” [P 21]). 

\hypertarget{undesirable-societal-impacts}{%
\subsubsection{Undesirable Societal Impacts}\label{undesirable-societal-impacts}}

26 scenarios (24.5\%) described undesirable impact on a societal level. Within the impact group, we distinguish between impacts on the economy, social cohesion, ethical use, religion, and the meta-physical level of humanness.

 Some scenario-writers anticipated a large-scale effect on the economic system. Scenarios described a future where chatbots replaced human jobs on a broader scale, which led to a rise in unemployment - and an even heavier competition regarding the remaining jobs. Unemployment was also often coupled with the influence of big tech that led to a concentration of power and a dependency on a few corporations (“I believe that people are making tech just to make it because it will sell or because it will help them sell something, not because it will help society in any way.” [P P8]).

 
Regarding social cohesion, one respondent outlined a declining birth rate due to the increasing isolation that users experience as a consequence of massive chatbot use; another respondent outlined a fostering of societal division due to the polarizing character of chatbots. Moreover, one respondent mentioned that the use of chatbots foster the abandonment of religion in society. Concerning the ethical component, some scenario-writers described a further acceleration of misinformation due to the growing capabilities of chatbots (“Considering how widespread "fake news" has been during the past few US elections I can only see that problem getting worse now that AI chatbots and AI image tools are more easily accessible to almost anyone.” [P P11]). In addition, some ethical aspects were mentioned only one or two times such as transparency issues, lack of fact-checking, and a loss of societal trust (“It's hard to have trust that the technology won't be abused by the people in power.” [P 35]). Lastly, some scenario-writers also reflected on a more metaphysical level what it means to be human in times of rapid technological development. The impacts here mostly described a discussion of “true” human domains that should not be impacted by chatbots (“I greatly value human connection and see it as a fundamentally important aspect of living, but what happens when the words of a human and the words of an AI are indistinguishable?” [P P9]). Thus, they amplified a return to human skills and a definition of domains where technology should not be applied (“The quality of production we have today is because people have spent their lives learning a craft and expanding on it with their own ideas. An AI can't do that, it can't create meaning out of what it makes” [P P7]).  

\hypertarget{predictors}{%
\subsubsection{Predictors of Impact Anticipations}\label{predictors}}

In a last step, we were interested in tracing the occurrence of desirable and undesirable impacts back to the sociodemographic information and AI related attitudes of the respondents. For this, we calculated two logistic regression models using (1) occurrence of desirable impacts in a scenario (2) occurrence of undesirable impacts in a scenario as dependent variables. We entered gender (1=male), race (1=White), age, educational level (1=Bachelor degree or higher), income, experienced discrimination, AI interest, media use of AI related news, and chatbot use as independent variables in the model (See Appendix \ref{appendix6} for full details).

Due to the relatively small sample size for quantitative analysis the model is underpowered and so the results here should be interpreted as exploratory. However, the models do explain 14.4 percent (M1) and 11.7 percent (M2) of the variance of the dependent variables. Turning to the predictors, we are able to detect one significant  association in the data and we also report on what appear to be some other patterns as well. While accounting for all of our control variables, the articulation of undesirable impacts of chatbots was found to be associated with identifying as White (p < 0.05), with 30 (59\%) of the group’s scenarios indicating undesirable impacts, in comparison to 24 (44\%) of non-White participant’s scenarios indicating undesirable impacts. Some other (non-statistically significant) patterns we see in the data are for race and the articulation of desirable impacts (18, 35\% for White, and 29, 53\% for non-White), and experienced discrimination and the articulation of desirable impacts (25, 51\% for experienced=1 or 2, and 22, 39\%, for experienced=3,4,5). 

\hypertarget{discussion}{%
\section{Discussion}\label{discussion}}

AI-based general purpose chatbots, such as ChatGPT, are part of the everyday private and professional lives of millions of users, and will be even more so in the future. As a general purpose technology without a concrete pre-defined purpose, personal chatbots could be used for a whole range of objectives, depending on the personal needs, contexts and tasks of an individual, and in doing so potentially impact a variety of values, people, and social contexts. By deploying scenario-writing at scale, we were able to map specific impact areas and distinguish between desirable as well as undesirable impacts of chatbots from a user-centric position, acknowledging also scenario-writers' social embedding, context and value expectations. This study thereby demonstrates that  scenario’s can be a fruitful instrument to actively engage users and tap into a multitude of diverse contexts and value expectations.

Our analysis showed that the anticipated impacts of general purpose chatbots vary considerably – and are two-sided. Users’ future anticipations of the impacts of chatbots are largely balanced between desirable and undesirable impacts. At the same time, we found that a larger proportion of undesirable impacts were described at the societal level than at the individual level, and we demonstrated that there are some tendencies in associations between sociodemographics and impact types supporting the need to engage broad and diverse sets of stakeholders when using scenario-based methods. As an anticipatory endeavor, this research suggests the need for a great deal of future work to increase the stock of knowledge and reduce the uncertainty around both the positive and negative potential impacts that were described. If the desirable futures that participants envision are to materialize, future research should assess the actual (rather than perceived) potential for improving well being when using chatbots, and that the performance characteristics of the technology do not hinder this potential. Likewise, undesirable futures should be met with assessments to determine the prevalence and severity of those potential negative outcomes, and to put in place mitigating measures through design, policy, or other responsibility assignments. In the following subsections we elaborate further on our study's findings. 

\hypertarget{individual-futures}{%
\subsection{Bright Individual Futures …}\label{individual-futures}}

One of the most frequently mentioned positive aspects of the use of advanced general purpose chatbots, is the anticipated impact on persons’ well-being and self-development. In these future anticipations, chatbots are perceived as more than a technical tool, but rather as a companion that helps people manage their lives as a work-life-balance coach, a dietary assistant, or even a therapist. Many scenario writers perceived chatbots as tools that people can use to strengthen their well-being, with characters that felt encouraged and found motivation for navigating through tough times and complying with their various tasks. Given the prevalence of the codes and the positive connotation of the scenarios, it seems plausible that users want to use chatbots for this purpose now and in the future.  

Further, many respondents expect chatbots to develop the strength and purpose that current chatbots already possess even further: fostering efficiency and being convenient for their use. Chatbots here are perceived as tools that contribute to the productivity of users and make life easier. Many scenario-writers expect that chatbots will fulfill these tasks satisfactorily. That is not overly surprising given to the capabilities these chatbots already have. However, these anticipated benefits are interesting, when we see them in relation to other impacts, for instance, benefits for work. Due to efficiency improvement and easy accessibility, some respondents outlined the flourishing of (small) business as the chatbots can take over routine tasks and enable business owners and employees to focus on other tasks. Interestingly, only a few scenarios outline positive effects on a societal level. While some aspects like the fostering of multiculturalism and a strengthening of collective action are mentioned, for the majority of scenario writers the truly added benefit seems to be in the personal sphere, and comes with costs for society.

\hypertarget{detrimental-effects}{%
\subsection{… and Detrimental Societal Side-Effects …}\label{detrimental-effects}}

When it comes to negative impacts on the individual level, some seem to be related to malfunctioning, but most are less related to chatbots themselves rather than what the consequences are of relying on technology. At the individual level these are: loss of skills, loss of creativity, loss of human interaction skills, in other words; what technology turns us into, coupled with a sense of loss of control and, related, growing dependency. On the societal level: many scenarios center here at the use of AI against users or society: by big tech, employers or governments to replace humans, pollute the public sphere or resulting in power imbalances while the loss of cohesion and declining trust disempower society.

The use of general purpose chatbots spurs a debate about what skills humans need, what creativity is, and if people lose their ability to think on their own, when chatbots can create texts, or provide information on basically every subject. This opens up a meta-physical debate about humanness in terms of AI proliferation \cite{guzman_artificial_2020}. This impact category goes beyond those identified in prior AI impacts assessments that are mostly concerned with tangible and measurable risks. However, especially in consideration of value-based approaches to AI development and implementation, it is crucial to also deal with underlying questions. Or, in the words of one respondent: “Artificial intelligence can be beneficial, but it also removes that face to face interaction that's important.” [P 36]. The negotiation on what tasks individuals and society want to pass to AI, and what should remain human requires a broad societal and ethical discussion that should include the moral values of individual members of society. However, that requires an engagement with society and cannot only be solved by top-down impact assessments or value-sensitive design \cite{crawford_atlas_2021, kieslich_commentary_2022, rahwan_society---loop_2018, zuger_ai_2023}. Furthermore, we found a frequent mention of undesirable anticipated impacts on work-related issues. Again, these are impacts that are not directly related to the technology itself, but the way it is deployed and used to the detriment of individuals and society. Negative impacts on social cohesion and human interaction increase societal vulnerability and the inability for collective action, for example in relation to large technology corporations or institutions. Given the capabilities of personal chatbots, many respondents fear the automation of jobs. Whereas this is not a new issue, scenario-writing creates vivid examples of users’ anticipation that make this issue more tangible. Scenarios like these can be used as a starting point in conversations with policymakers or companies that are faced with the task of mitigating harms and implementing AI in an employee-friendly way. The economic impacts on jobs was also transferred to the societal level as multiple respondents were concerned about a rise in unemployment. Indeed, the negative impact of general purpose chatbots on employment is a widely discussed issue in academia and the public discourse. Some of the scenario-writers seem to pick up on this and enrich these trends with their own thoughts and experiences. In the value questions, respondents reflected about the large-scale effects on employment and outlined the detrimental effects that a widespread use of chatbots could have. 

The overall picture that emerges is that chatbots are widely expected and welcomed as useful tools to advance individual personal goals, but that they come with considerable concerns about the more medium to long term implications for human skills and qualities, and the (societal) backlash of being overly reliant and dependent on technology. The results seem to point to the conundrum that individuals see a lot of potential value for themselves, but see potential undesirable consequences in society. This means individuals will want to use the technology, but the open question then is: who invests in making sure the negative social detriments are addressed, and that the social desirable outcomes are invested in further too. Interestingly, many concerns do not center so much around the technology and the way it is being designed, but rather the way it is being deployed and relied on, as well as a lack of control in the face of power imbalances and growing dependencies. Findings like these trigger important questions for current policy and regulatory approaches towards AI, such as the European AI Act. These approaches focus in the first place on making the technology itself safer, more responsible so that it can be trusted and safely used. The majority of our respondents, however, were not even concerned about the lack of safety or malfunctioning of the technology itself, but the broader societal implications, suggesting to truly address individual’s concerns it is not enough to make the technology itself safer and trustworthy, but its development and deployment must come with investment in human skills, continued learning and developing and a greater sense of individual and democratic control over its implementation. 

\hypertarget{long-tail}{%
\subsection{… and the Long-Tail of Impacts.}\label{long-tail}}

Respondents identified various undesirable impacts that were grounded in their perceptions and experiences with the technology. We stress that we are not only interested in the most frequently occurring issues, but also in the long-tail of other identified impacts which is elucidated by the anticipatory approach \cite{amosbinks_anticipatory_2023}. For instance, AI was described by one participant as a threat to religion. Other respondents described the value of chatbots for specific subgroups with mental issues. For instance, one respondent wrote: “I feel that chatbots and AI can be incredibly helpful to people that need guidance in their day-to-day life activities. [...] there are many areas where AI can help improve and lend a helping hand to improve the lives of those who are busy or struggle for other reasons such as ADHD.” [P 4]. Here, it is especially important that future research dive deeper into the potential benefits and risks for these groups and assess how personal chatbots affect these groups. One way to address this is in conducting inclusive workshops with these groups to map out their ideas and perceptions in far greater detail. Citizen-centered scenario-writing enables researchers to detect those potential impacts and set a reference point for more targeted impact assessments. For instance, to flesh out the long tail, faith-based groups could be engaged to better understand how chatbots could impact religious practices, or patients might be engaged to elaborate on specific use cases that may be meaningful in their own mental health. 

\hypertarget{cognitive-diversity}{%
\subsection{Cognitive Diversity Matters}\label{cognitive-diversity}}

A major benefit of a citizen-centric large-scale scenario-writing exercise is that we can enrich cognitive diversity \cite{brey_ethics_2017, nikolova_rise_2014} and, thus, expand the current AI impact assessment landscape. While we already reported on some interesting findings on the individual scenario level, we now turn to a reflection about quantitative patterns and trends in the anticipation of impacts. Due to the relatively low case number (for a quantitative analysis), we stress that we only can detect trends, but also that these trends indicate possible directions for more highly powered follow-up research. We found a trend that people, who experienced discrimination at least sometimes, tend to be more concerned about the future use of personal chatbots: they less frequently outline desirable impacts. This is plausible as the discriminatory impact of AI tools is well-established in previous literature \cite{chiusi_automating_2020, crawford_atlas_2021}. Former public opinion studies have shown that these groups tend to engage less with the issue of AI at all \cite{jakesch_how_2022, kieslich_ever_2023}, however our findings indicate that the detrimental effects for people who experienced discrimination are salient to  this group. It may be that framing this exercise as scenario-writing (rather than a poll) was an approachable way for groups to express such issues and provide an avenue to give a voice to these individuals and groups. To our surprise, we found that people of color in comparison to White respondents associated personal chatbots with desirable impacts for individuals, while White respondents articulated more undesirable impacts, and that this relationship was apparent after controlling for education, income, and AI interest. Given the discriminatory impact that AI technology can have on people of color, this finding is rather surprising and warrants further investigation, such as through a qualitative interview study that could unpack the nature of such a racial disparity. 

\hypertarget{limitations}{%
\subsection{Limitations}\label{limitations}}

As with most studies that rely on sampling through online access panels, we are not able to reach all groups that can be affected by the use of chatbots. For this study, that especially encompasses people with low digital literacy skills who have no access to online access panels. Additionally, we only included fluent English speakers residing in the US in our sample. While pragmatic, these choices leave out some groups in the data collection. Thus, we acknowledge the need to collect data from people who can't be reached by standard survey research in future work. This can be done, for instance, by following examples from the FAccT community \cite{bray_radical_2022, dillahunt_eliciting_2023, helm_diversity_2022, lee_webuildai_2019} to have face-to-face workshops with vulnerable groups, for instance, in cooperation with non-profit organisations. We highly encourage future work to follow this path to better augment the current debate about the impacts of AI-based general purpose chatbots.


\hypertarget{conclusion}{%
\section{Conclusion: Risk Mitigation and Benefits Amplification}\label{conclusion}}

General purpose AI chatbots pose the challenge that they can be used for a variety of tasks and can lead to numerous impacts -- also depending on the purpose of the application and the entity using the general purpose chatbots. This makes it inherently difficult to anticipate the impact of this technology, and also to mitigate harms or amplify beneficial uses. In our study, we present a method for mapping the impacts of such a general-purpose technology. We show that citizen-centered scenario writing can serve as a tool to engage individuals and add a user-centered perspective to impact assessments. These impact descriptions, unlike other impact assessments, are grounded in the reality of citizens and potential users of the technology, and thus can make explicit what these impacts might look like, as well as highlight potential value conflicts that arise when people interact with general-purpose chatbots. Our approach can contribute to advancing impact assessment methodologies by (1) actively engaging individual users from a diversity of backgrounds and perspectives, (2) qualitatively exploring public values dimensions by explaining why participants feel positive or negative about a particular scenario, and (3) mapping adoption areas and potential values at stake, which in turn can inform the need for more targeted impact assessments and inform policy makers, developers, and deployers about the expectations and concerns of users in a specific scenario. 

We mapped the desired and undesired impacts that our respondents considered most likely or found most worrisome or hopeful from their individual perspectives. Not all of these scenarios may be realistic, but the anticipations, hopes, and fears of citizens are. Unlike impact assessment frameworks that must identify the real impact of a particular technology application on (particular) values, the added value of this study is that it can provide guidance on where to look first. This is particularly important for general purpose technologies that can be potentially used for so many different purposes. It is important for regulators, scholars, and companies to see and address these potential impacts in order to work towards social well-being. While undesirable scenarios may point to risks or value conflicts that future risk assessments or policy measures might have to address, the desirable scenarios can be a starting point for discussions of what is needed to unlock the potential of chatbots for individual users as well as society. 

In the end, future approaches to AI should find a way to balance both desirable and undesirable impacts. This is also articulated by some respondents of our study, who reflected about a responsible and balanced approach to the future of AI development and implementation: “The impacts [...] emphasize the balance between technological advancement and human judgment. Maintaining this balance is essential for moral decision-making and protecting human intuition in an increasingly AI-dependent world.” [P1]. Given the pace of technology development it is of great importance to investigate and develop responsible approaches for how society can and should deal with the rise of general purpose systems. Scenario-writing offers a useful starting point for this matter.  

\hypertarget{ethics-statement}{%
\section*{Ethics Statement}\label{ethics-statement}}

In conducting this research, we adhered to the standards of ethical integrity and responsibility for research with human participants. We ensured that all participants were fully informed about the nature and purpose of the research and provided their informed consent. Participant confidentiality and data privacy were rigorously maintained throughout the study. Ethical guidelines, including those pertaining to non-discrimination, fairness, and respect for individuals, were strictly followed. The research methods were designed to minimize potential harm or discomfort to participants. Any conflicts of interest were disclosed and managed
appropriately. We received ethical approval from the ethics board of our university.

\hypertarget{conflict-interest}{%
\section*{Conflicts of Interest}\label{conflict-interest}}

The authors declare that they have no known competing financial interests or personal relationships that could have appeared to influence the work reported in this paper.

\begin{acks}

The funding for this research was provided by UL Research Institutes through the Center for Advancing Safety of Machine
Intelligence.

\end{acks}

\bibliographystyle{ACM-Reference-Format}
\bibliography{CASMI_camera_ready}

\newpage

\hypertarget{appendix}{%
\section*{Appendix}\label{appendix}}

\hypertarget{appendix1}{%
\subsection*{Appendix 1: Item wording}\label{appendix1}}

\textbf{Experienced discrimination.} Experienced discrimination was measured with five items adapted from Williams and colleagues adopted from \cite{williams_racial_1997}: Respondents indicated on a five-point scale how often the following things happened to them in everyday life (1=never, 2=seldom, 3=sometimes, 4=often, 5=very often): Someone acts as if they are afraid of you; You receive worse service than other people in restaurants or stores.; You are threatened or harassed.; Someone acts as if you are not taken seriously.; You are treated with less respect than other people. We calculated a maximum score for each respondent indicating the maximum value of a respondent score on one of the five items (M=2.68; SD=0.94).

\textbf{AI interest.} We measured interest in AI with four items (I follow developments related to artificial intelligence with great curiosity; In general, I am very interested in artificial intelligence; I read articles about artificial intelligence with great attention; I watch or listen to articles about artificial intelligence with great interest) on a five point scale (1=does not apply at all; 5=fully applies) adopted from Došenović et al. \cite{dosenovic_methodensteckbrief_2022}. We calculated a mean index for interest in AI across the items (Cronbachs \(\alpha\)=.91; M=3.09; SD=0.92). 

\textbf{Media consumption of AI-related news.} We measured media consumption of AI with four items adapted from Došenović et al. \cite{dosenovic_methodensteckbrief_2022}, asking respondents: “Thinking back over the last month, how often have you…”. on a six point scale (1=not at all; 2=once a month; 3=about two or three times a month; 4=about once a week; 5=several times a week; 6=daily). The items read as follows: (1) read about AI in a daily or weekly newspaper or magazine - including the corresponding online editions?; (2) read about AI in blogs or online-only magazines?; (3) watched news on AI on television?; (4) read about AI on social media? We calculated a maximum score for each respondent indicating the maximum value of a respondent score on one of the four items (M=4.06; SD=1.26). 

\textbf{Personal use of chatbots.} We measured personal use of AI-based chatbots with a self-developed item (How often do you use personal chatbots? With personal chatbots we mean AI-based technological applications with which you can engage in conversations and/or which carry out tasks as personal assistants. An example for a personal chatbot is ChatGPT.) on a six point scale (1=not at all; 2=once a month; 3=about two or three times a month; 4=about once a week; 5=several times a week; 6=daily).

\newpage
\hypertarget{appendix2}{%
\subsection*{Appendix 2: Task Description}\label{appendix2}}

\textbf{Please read the following instructions here and on the next pages very carefully as you need this information to perform your writing task.}

In this exercise we ask you to write a short (~ 300 words) fictional scenario to explore how the use of AI-based personal chatbots could create \textbf{impacts for users and/or society} in five years from now.

A \textbf{scenario} is a short \textbf{fictional story} that includes: 1) a \textbf{setting} of time and place, 2) \textbf{characters} with particular motivations and goals, and 3) a \textbf{plot} that includes character actions and \textbf{events} that lead to some impact of interest.

\textbf{AI-based personal chatbots} can be used to hold conversations with users and/or to perform tasks as a digital assistant. They are not designed for a specific topic, but are used to answer a wide range of requests.

Specific Instructions:
\begin{itemize}
\item Please develop your scenario \textbf{based on your own perspectives, experiences, and knowledge}.
\item Please develop your scenario to be \textbf{creative and original} in regard to the \textbf{impacts of personal chatbots for users and/or society}.
\item Please choose your \textbf{setting} to be five years in the future and in the society where you currently reside.
\end{itemize}
On the next pages you will see more details on the technology for the scenario we ask you to write. Please take your time to familiarize yourself with the information.

\newpage
\hypertarget{appendix3}{%
\subsection*{Appendix 3: Technology Description}\label{appendix3}}

\textbf{Technology}

An AI-based personal chatbot is a digital assistant that can be used to hold conversations with users and/or to perform tasks as a digital assistant. These chatbots are able to process and respond to text or voice inquiries in a conversational way, mimicking human interaction.

\textbf{Capabilities}
\begin{itemize}
\item Personal chatbots can be controlled using text- or voice-based prompts which provide task instructions and input data. For instance you can prompt it with: 	
\begin{itemize}
\item "Summarize the following text: <text>"
\item "Translate the following text into English: <text>"
\item "Explain <issue> in easy to understand language"
\item "Schedule my appointment about <issue>"
\item "Provide me with information about <issue>"	
\end{itemize}
\end{itemize}
\begin{itemize}
\item Personal chatbots can be personalized so that they can tailor information to different user preferences based on prior user interactions.
\item Personal chatbots are not designed for a specific topic, but are used to answer a wide range of requests.
\item Personal chatbots can be used to schedule appointments, search for information, send messages and summarize, personalize or translate text. They are set up as conversational agents that end-users can interactively communicate with, and can be incorporated into other technologies like search engines, social media, smart home devices, or mail programs. 	 	
\end{itemize}

\textbf{Limitations}

\begin{itemize}
\item \textbf{Accuracy}: This technology does not always output text that is accurate.
\item \textbf{Privacy and Security}: This technology processes sensitive personal information of users.
\item \textbf{Biases}: The outputs from this technology can be biased based on the data used to train the system, which typically reflects common societal biases (e.g. racial or gender).
\item \textbf{Lack of Contextual Understanding}: The technology may not understand the context of a question and/or user's intentions.
\end{itemize}
Here you can see a screenshot of the interface of a personal chatbot. Users can type in their question/task in the chatbox and the chatbot will provide the answer as a reply.

\newpage
\hypertarget{appendix4}{%
\subsection*{Appendix 4: Screenshot of an AI-based chatbot}\label{appendix4}}

\begin{figure*}[h]
    \centering
    \includegraphics[width=0.95\linewidth]{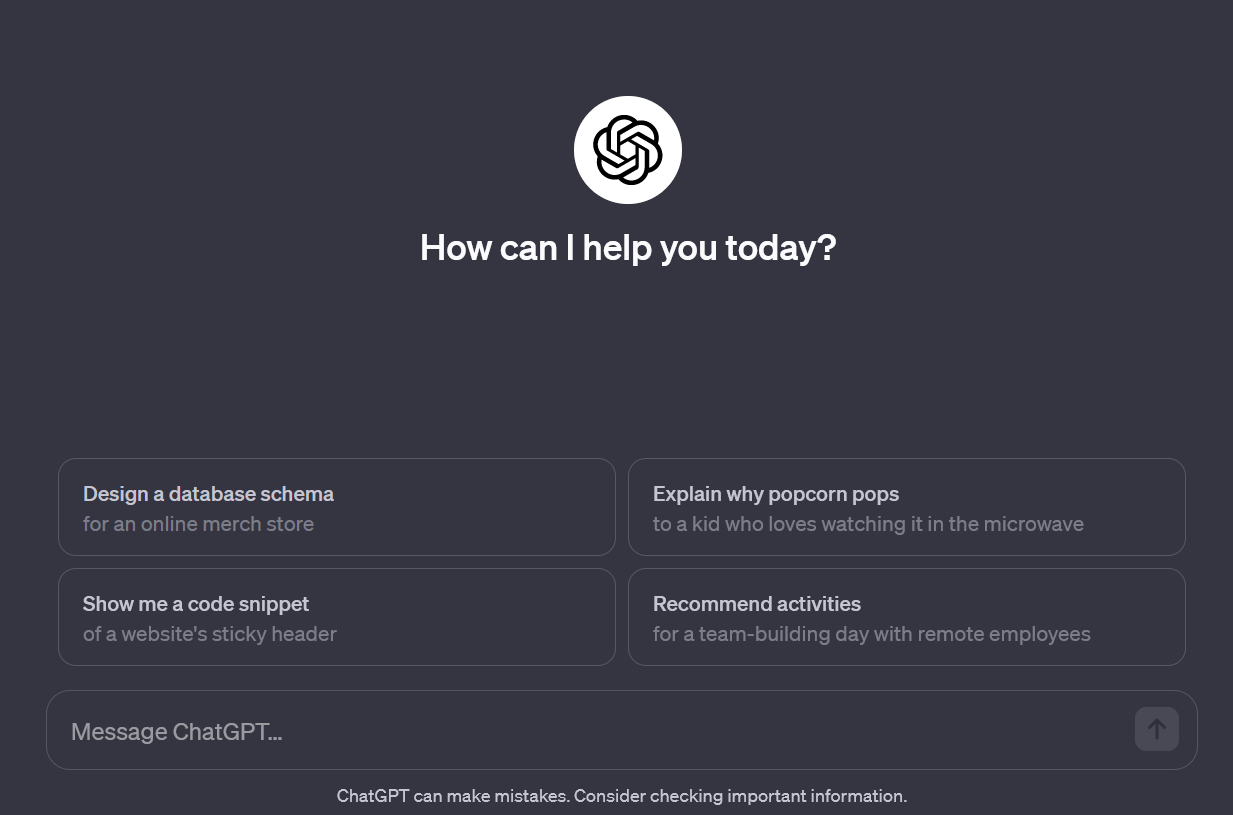} 
    \caption{Screenshot of ChatGPT}
    \label{fig:yourlabel}
\end{figure*}

\newpage
\hypertarget{appendix5}{%
\subsection*{Appendix 5: Sample Description}\label{appendix5}}

For data collection we relied on the participant pool of the market research provider Prolific. We aimed for a final sample size of approximately 100 respondents as we i) aimed for a broad mapping of potential impacts of chatbots, and ii) wanted to conduct a quantitative data analysis on the occurrence of desirable and undesirable future impacts. The minimum required sample size for a regression model is ten respondents per independent variable. As our models include nine independent variables, a finale sample of approximately 100 respondents satisfies this criteria. However, we acknowledge that sampling at the minimum sample size leads to the possibility that the model is still under-powered.  

53 respondents self-identified as male, 50 as female and 3 as non-binary. The average age of respondents is 36.81 years (SD=13.54). Education level is high with 24 respondents holding a graduate degree, 35 a Bachelor degree, 15 an associate degree, and 20 some college and 12 with a high-school degree. Regarding race, 51 respondents identified as White, 20 as Black or African American, 14 as Asian or Pacific Islander, 11 as Hispanic, 9 as multiple Ethnicity and 1 as American Indian or Alaskan native. Household income is distributed as follows: 13 respondents reported having an household income between, 0 and \$19,999 USD, 20 between \$20,000 USD and \$49,999 USD, 37 between \$50,000 USD and \$89,99 USD, 20 between \$90,000 USD and \$129,999 USD, 7 between \$130,000 USD and \$149,000 USD, and 6 over \$150,000 USD (3 preferred not to answer).   

\newpage
\hypertarget{appendix6}{%
\subsection*{Appendix 6: Logistic Regressions on 1) Desirable Impacts and 2) Undesirable Impacts}\label{appendix6}}

\begin{table*}[h]
\caption{Logistic Regressions on 1) Desirable Impacts and 2) Undesirable Impacts}
\centering
\resizebox{\linewidth}{!}{
\begin{tabular}[t]{lllllllllllll}
\toprule
\multicolumn{1}{c}{ } & \multicolumn{6}{c}{Model 1} & \multicolumn{6}{c}{Model 2} \\
\cmidrule(l{3pt}r{3pt}){2-7} \cmidrule(l{3pt}r{3pt}){8-13}
\multicolumn{3}{c}{ } & \multicolumn{3}{c}{95\% CI for Odds Ratio} & \multicolumn{3}{c}{ } & \multicolumn{3}{c}{95\% CI for Odds Ratio} \\
\cmidrule(l{3pt}r{3pt}){4-6} \cmidrule(l{3pt}r{3pt}){10-12}
  & b & SE & Lower & Odds Ratio & Upper & p & b & SE & Lower & Odds Ratio & Upper & p\\
\midrule
Intercept & -1.181 & 1.195 & 0.028 & 0.307 & 3.148 & 0.323 & -0.915 & 1.190 & 0.037 & 0.401 & 4.066 & 0.442\\
Gender (1=male) & 0.016 & 0.475 & 0.398 & 1.016 & 2.591 & 0.973 & 0.086 & 0.463 & 0.437 & 1.090 & 2.719 & 0.853\\
Race (1=White) & -0.621 & 0.457 & 0.216 & 0.538 & 1.310 & 0.175 & 0.988 & 0.464 & 1.101 & 2.685 & 6.873 & 0.033\\
Age & 0.024 & 0.018 & 0.989 & 1.025 & 1.063 & 0.187 & -0.015 & 0.018 & 0.950 & 0.985 & 1.020 & 0.391\\
Educational Level (1=high) & 0.267 & 0.507 & 0.483 & 1.305 & 3.570 & 0.599 & -0.455 & 0.502 & 0.233 & 0.634 & 1.690 & 0.364\\
Income & 0.031 & 0.183 & 0.719 & 1.032 & 1.483 & 0.864 & -0.015 & 0.178 & 0.692 & 0.985 & 1.401 & 0.931\\
Experienced Discrimination & -0.382 & 0.246 & 0.414 & 0.682 & 1.096 & 0.121 & 0.242 & 0.237 & 0.804 & 1.273 & 2.050 & 0.308\\
AI Interest & 0.215 & 0.304 & 0.684 & 1.239 & 2.276 & 0.480 & 0.221 & 0.304 & 0.688 & 1.247 & 2.293 & 0.467\\
Media Use AI & 0.205 & 0.200 & 0.829 & 1.228 & 1.834 & 0.306 & -0.039 & 0.194 & 0.654 & 0.962 & 1.413 & 0.840\\
Chatbot Use & -0.107 & 0.164 & 0.647 & 0.898 & 1.235 & 0.513 & 0.040 & 0.161 & 0.758 & 1.041 & 1.434 & 0.805\\
\bottomrule
\multicolumn{13}{l}{\rule{0pt}{1em}\textit{Note: }}\\
\multicolumn{13}{l}{\rule{0pt}{1em}Model 1: Nagelkerke $R^2$=.144, Model $\chi^2$(9)=11.32, p=.255, p=.243; Model 2: Nagelkerke $R^2$=.117, Model $\chi^2$(9)=9.07, p=.43}\\
\end{tabular}}
\end{table*}

\end{document}